\documentstyle[aps,twocolumn]{revtex}

\begin{document}

\title{Comment on ''Photonic Band Gaps: Noncommuting Limits and the 'Acoustic
Band'\,"}
\maketitle
This Comment concerns the low-frequency dielectric constant $\epsilon _{eff}$
of 2D photonic crystals (PC's), constituted of circular cylinders (in
vacuum) characterized by an {\it {extremely high or infinite value of the
dielectric constant}}$\epsilon .$ In Ref.\cite{one} and in two related
papers \cite{two,three} it was claimed that the quasistatic approach, based
on the determination of the slope of the "acoustic" photonic band, leads to a
value of $\epsilon _{eff}$ that is substantially smaller than the $\epsilon
_{eff}$ obtained by an electrostatic method of ''homogenization''. Here we point out
that this conclusion is incompatible with Electrodynamics and demonstrate that the two approaches do give the same  results, essentially (within computational error).

Recently
we have reported general formulas for the principal dielectric constants of
an arbitrary 2D PC\cite{four}. These formulas are obtained by taking the limit $\omega, k \rightarrow 0$ directly in the wave equation for periodic media. For isotropy in the plane of periodicity $%
A_{xy}=0$ and $A_{xx}=A_{yy},$ so that $\epsilon _1=\epsilon _2 \equiv \epsilon
_{eff}.$ Also, for $\epsilon \rightarrow \infty, $ $\overline{\eta }=1-f$ and $%
\eta ({\bf{G}})=-2fJ_1(Ga)/Ga$, where $f$ is the filling fraction of the
cylinders (radius $a$), $J_1$ is a Bessel function, and $\bf{G}$ is
the reciprocal lattice vector. Then, for any isotropic 2D lattice, it follows 
from Ref.\cite{four} that

\begin{eqnarray}
\epsilon _{eff}&=&1/\left\{ 1-f+f\sum_{\bf{G,G}^{\prime }\neq 0}{\bf{
G\cdot G}^{\prime }}\frac{J_1(Ga)}{Ga}\frac{J_1(G^{\prime }a)}{G^{\prime }a}
\right. 
\nonumber \\ \cr
&\times&\left.\left[{ \bf{G\cdot G}^{\prime }}\frac{J_1(\left|{ \bf{G-G}^{\prime
}}\right| a)}{\left| {\bf{G-G}^{\prime }}\right| a}\right] ^{-1}\right\} ,
\label{one}
\end{eqnarray}
where the exponent ''$-1$'' implies matrix inversion. 

We
have applied Eq. (1) to a square lattice; the last column of Table I gives
the corresponding index of refraction $N_{eff}=\sqrt{\epsilon _{eff}}$.
We see that there is excellent agreement with the static calculations \cite
{five}, however poor accord with the results deduced from the slope of the
acoustic band\cite{one,two}. The authors of Ref.\cite{one} attribute the discrepancy between the 2nd and
3rd columns to noncommuting limits $\epsilon \rightarrow \infty $
and $k\rightarrow 0$ ($k$ is the Bloch vector). However, in our approach, 
these limits  {\it {do commute}}, i.e. Eq. (1) and the results in the 4th 
column of the Table I are {\it independent on the order of taking the limits}.
Precisely the commutation of these two limits
\begin{table}[h]
\caption{Effective indeces of refraction $N_{eff}$ for three \\
values 
of the filling fraction $f$.}
\begin{tabular}{|c|c|c|c|}  
                           & quasistatic          & static             & quasistatic \\ 
{\it f}                    & approach,           & approach,      & approach, \\
                           & Refs.  \cite{one,two}          & Ref. \cite{five}         & this Comment \\ \hline
                                  
0.363                     & 1.1736           & 1.4663          & 1.4661 \\ 
0.454                     & 1.2151           & 1.6435          & 1.6443 \\
0.554                     & 1.2782           & 1.9135          & 1.9129  \\   
\end{tabular}
\end{table}
\noindent provides the possibility to obtain the solution of an electrostatic problem for conductors from the solution for dielectics bodies of the same shape by taking the limit $\epsilon\rightarrow \infty$\cite{six}. We believe that the quasistatic
results of Ref.\cite{one,two,three} are erroneous for very large and for
infinite values of $\epsilon $, and that the reason for this resides in the
truncation of the generalized Rayleigh identity to finite values of the
order $l$ ($l\leq 20$). If this truncation is performed {\it {prior to
taking the limit }$\epsilon \rightarrow \infty $}, the stability of the
numerical results does {\it not} guarantee that they are correct. (Indeed in Eq. (%
\ref{one}) the truncation of the ${\bf G}$ values is the last step).We expect that
for $\epsilon =10^{10}$ \cite{one}, $l_{\max }$ should be extremely large and that
for $\epsilon =\infty $ also $l_{\max }\rightarrow \infty $. This is based on the fact
that, the higher the dielectric contrast between the constituents, the greater the computational effort required.
On the other hand, in Ref.\cite{four} the contrast is given in terms of the
{\it reciprocal} dielectric constants, leading to excellent convergence. 

The physical explanation of the disagreement \cite{one,two,three} has been
attributed to the extremely small wavelength within the
cylinders. However, this argument is invalid, because the homogenization (in periodic systems) 
implies averaging over scales much larger then the {\it Bloch wavelength} $2\pi/k$. Local wavelength in the material is meaningless.
The reassuring conclusion is
that Electrostatics is still the $\omega \to 0$
limit of Electrodynamics - even
for PC's composed of cylinders with an infinite dielectric constant.

\vspace{.3cm}
P.Halevi$^{1}$, A.A. Krokhin$^{2}$, and J. Arriaga$^{2}$

\indent $^{1}$Instituto Nacional de Astrof\'\i sica, Optica y \\
\indent  Electr\'onica, Apdo. Post. 51, Puebla, 72000, M\'exico \\
\indent $^{2}$Instituto de F\'\i sica, Universidad Aut\'onoma de Puebla, \\
\indent Apdo. Post. J-48, Puebla, 72570, M\'exico  

\vspace{0.3cm}
\noindent Received  21 September 1999 

\end{document}